\documentclass[sigconf,authorversion,nonacm]{acmart}
\usepackage{amsmath,amsfonts}
\usepackage{algorithmic}
\usepackage{algorithm}
\usepackage{array}
\usepackage{textcomp}
\usepackage{stfloats}
\usepackage{url}
\usepackage{verbatim}
\usepackage{graphicx}
\usepackage{enumitem}
\usepackage[belowskip=1pt,aboveskip=1pt]{caption}
\usepackage[font=small,labelfont=sf,textfont=sf]{subcaption}
\usepackage[all]{xy}
\usepackage{rotating}
\usepackage{multirow}
\usepackage{booktabs}
\usepackage{quantikz}
\usepackage{adjustbox}
\usepackage{float}
\usepackage{tabularray}
\usepackage[nodisplayskipstretch]{setspace}
\usepackage{makecell}
\usepackage{tikz}
\usepackage{quantikz}
\usepackage{amsmath}

\newcommand{\rowSep}{0.22cm}
\newcommand{\colSep}{0.18cm}

\newcommand{\myCnot}{
\begin{quantikz} [row sep= \rowSep, column sep= \colSep]
	\lstick{$a$} & \ctrl{1} & \qw & \rstick{$a$}\\
	\lstick{$b$} & \targ{}  & \qw & \rstick{$a \oplus b$}
\end{quantikz}
}

\newcommand{\myToffoli}{
\begin{quantikz} [row sep= \rowSep, column sep= \colSep]
	\lstick{$a$} & \ctrl{1} & \qw & \rstick{$a$} \\
	\lstick{$b$} & \ctrl{1} & \qw & \rstick{$b$} \\
	\lstick{$c$} & \targ{}  & \qw & \rstick{$a \cdot b \oplus c$}
\end{quantikz}
}

\newcommand{\xorNoCarryTwo}{
	\begin{quantikz} [row sep= \rowSep, column sep= \colSep]
		\lstick{$a_0$} & \targ{} & \qw & \rstick{$s_0$} \\
		\lstick{$b_0$} & \ctrl{-1} & \qw & \rstick{$b_0$} \\
		\lstick{$a_1$} & \targ{} & \qw & \rstick{$s_1$} \\
		\lstick{$b_1$} & \ctrl{-1} & \qw & \rstick{$b_1$} \\
	\end{quantikz}
}

\newcommand{\xorBCarryTwo}{
	\begin{quantikz} [row sep= \rowSep]
		\lstick{$a_0$} & \targ{} 	& \qw & \rstick{$s_0$} \\
		\lstick{$b_0$} & \ctrl{-1} 	& \qw & \rstick{$b_0$} \\
		\lstick{$a_1$} & \targ{} 	& \qw & \rstick{$s_1$} \\
		\lstick{$b_1$} & \ctrl{-1} 	& \qw & \rstick{$s_2$} \\
	\end{quantikz}
}

\newcommand{\xorToffoliCarryTwo}{
	\begin{tikzcd} [row sep= \rowSep, column sep= \colSep, ampersand replacement=\&]
		\lstick{$a_0$} \& \targ{}   	\& \qw 			\& \qw \& 		\rstick{$s_0$} \\
		\lstick{$b_0$} \& \ctrl{-1} 	\& \qw 			\& \qw \& 		\rstick{$b_0$} \\
		\lstick{$a_1$} \& \ctrl{1} 		\& \targ{}		\& \qw \& 	 	\rstick{$s_1$} \\
		\lstick{$b_1$} \& \ctrl{1} 		\& \ctrl{-1}	\& \qw \& 		\rstick{$b_1$} \\
		\lstick{$z$}   \& \targ{} 		\& \qw   		\& \qw \& 		\rstick{$z \oplus s_2$}
	\end{tikzcd}
}

\newcommand{\passNoCarryTwo}{
	\begin{quantikz} [row sep=0.48cm]
		\lstick{$a_0$} & \qw & \rstick{$s_0$} \\
		\lstick{$b_0$} & \qw & \rstick{$b_0$} \\
		\lstick{$a_1$} & \qw & \rstick{$s_1$} \\
		\lstick{$b_1$} & \qw & \rstick{$b_1$} \\
	\end{quantikz}
}

\newcommand{\passBCarryTwo}{
	\begin{quantikz} [row sep=0.48cm]
		\lstick{$a_0$} & \qw & \qw & \rstick{$s_0$} \\
		\lstick{$b_0$} & \qw & \qw & \rstick{$b_0$} \\
		\lstick{$a_1$} & \qw & \qw & \rstick{$s_1$} \\
		\lstick{$b_1$} & \qw & \qw & \rstick{$s_2$} \\
	\end{quantikz}
}

\newcommand{\orcidnew}[1]{\href{https://orcid.org/#1}{\includegraphics[width=8pt]{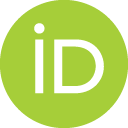}}}

\AtBeginDocument{%
  \providecommand\BibTeX{{%
    \normalfont B\kern-0.5em{\scshape i\kern-0.25em b}\kern-0.8em\TeX}}}

\setcopyright{none} 
\begin{document}

\title{Noise-Resilient and Reduced Depth\\ Approximate Adders for NISQ Quantum Computing}


\author{Bhaskar Gaur \orcidnew{0000-0001-6738-6890}}
\affiliation{%
	\institution{University of Tennessee}
	\city{Knoxville}
	\state{TN}
	\country{USA}}
\email{bgaur@vols.utk.edu}

\author{Travis S. Humble \orcidnew{0000-0002-9449-0498}}
\affiliation{%
	\institution{Oak Ridge National Laboratory}
	\city{Oak Ridge}
	\state{TN}
	\country{USA}}
\email{humblets@ornl.gov}

\author{Himanshu Thapliyal \orcidnew{0000-0001-9157-4517}}
\affiliation{%
	\institution{University of Tennessee}
	\city{Knoxville}
	\state{TN}
	\country{USA}}
\email{hthapliyal@utk.edu}

\renewcommand{\shortauthors}{Bhaskar Gaur, Travis S. Humble, \& Himanshu Thapliyal} 


\vspace{5mm}
\begin{abstract}
The "Noisy intermediate-scale quantum" NISQ machine era primarily focuses on mitigating noise, controlling errors, and executing high-fidelity operations, hence requiring shallow circuit depth and noise robustness. Approximate computing is a novel computing paradigm that produces imprecise results by relaxing the need for fully precise output for error-tolerant applications including multimedia, data mining, and image processing.  We investigate how approximate computing can improve the noise resilience of quantum adder circuits in NISQ quantum computing. We propose five designs of approximate quantum adders to reduce depth while making them noise-resilient, in which three designs are with carryout, while two are without carryout. We have used novel design approaches that include approximating the Sum only from the inputs (pass-through designs) and having zero depth, as they need no quantum gates. The second design style uses a single CNOT gate to approximate the SUM with a constant depth of O(1). We performed our experimentation on IBM Qiskit on noise models including thermal, depolarizing, amplitude damping, phase damping, and bitflip: (i) Compared to exact quantum ripple carry adder without carryout the proposed approximate adders without carryout have improved fidelity ranging from 8.34\% to 219.22\%, and (ii) Compared to exact quantum ripple carry adder with carryout the proposed approximate adders with carryout have improved fidelity ranging from 8.23\% to 371\%. Further, the proposed approximate quantum adders are evaluated in terms of various error metrics.
\end{abstract}


\keywords{approximate computing, noise, quantum adder, quantum circuit, quantum computing, NISQ, FTQ}

\maketitle

\section*{Acknowledgement}
This research used resources of the Oak Ridge Leadership Computing Facility, which is a DOE Office of Science User Facility supported under Contract DE-AC05-00OR22725.

\section{Introduction}
Quantum circuits serve as the basis for constructing quantum algorithms, making them helpful in applications such as quantum cryptography, image processing, and other scientific computations such as variational quantum algorithms for nonlinear problems, simulation of many-body systems, linear systems of equations HHL, singular value thresholding or triangle finding \cite{shor1994algorithms, xia2020design, lee2019hybrid, lubasch2020variational}. Researchers have used dedicated libraries of quantum arithmetic circuits in quantum programming languages such as Qiskit and Microsoft Quantum Development Kit, helping design resource-efficient quantum algorithm circuits \cite{MicrosoftSoftware, qiskitModel}. Further, quantum arithmetic circuits such as adders are used as benchmark circuits in the error-aware compilation in IBM’s 20-Qubit machines \cite{ibm20qubit}. The noisy intermediate-scale quantum (NISQ) era quantum hardware available today is unsuitable for prolonged operations due to error accumulation \cite{jayashankar2022achieving}. As a result, efforts are concentrated on managing noise, minimizing errors, and performing high-fidelity operations by developing quantum circuits with shallow circuit depth and noise resilience. However, the currently existing quantum adders are based on the ripple carry mechanism in which the final carry depends on a long carry path that inflates the circuit depth and gate count leading to a proportional rise in noise \cite{cuccaro2004new, thapliyal2013design}. Exact quantum adders by Cuccaro et al. \cite{cuccaro2004new}, referred to as CQA1 with output carry and CQA0 without output carry, and TPL13 \cite{thapliyal2013design}, an optimized quantum ripple carry adder with output carry by Thapliyal et al. are widely used designs of quantum ripple carry adders. Hence, we evaluate our proposed approximate adders against these designs to facilitate comparison with previous works.

Approximate computing (AC) is a novel computing paradigm that produces imprecise results by relaxing the need for entirely accurate or completely deterministic operations. Therefore, error-tolerant applications like data mining, multimedia, and image processing, often do not require fully accurate results, as imprecise or less-than-optimal results suffice. In the existing literature \cite{sajadimanesh2022practical}, approximate quantum multipliers are designed based on controlled quantum adders that are focused on optimizing depth and T-count. However, the designs of quantum adders perform controlled additions, therefore, have resource overhead compared to non-controlled quantum addition. Further, it utilizes an uncomputation gate that requires measurement operations, limiting its applicability to the NISQ domain due to increased noise susceptibility due to SPAM errors \cite{nielsen2002quantum}. Therefore, to the best of our knowledge, no existing designs of uncontrolled (normal) quantum addition use approximate computing applicable to the NISQ domain. Hence, we propose five new designs of quantum approximate adders that have improved noise resilience and reduced circuit depth. This is achieved by removing the carry propagation dependency during the addition operation by applying the principle of approximate computing. Since there is no carry dependency, each addition of input qubits operates in parallel, resulting in reduced logic gates and depth, thereby increasing noise resilience compared to exact designs of quantum adder circuits. To summarize, we make the following contributions:
\vspace{1mm}

\begin{itemize} [leftmargin=+.2cm, topsep=4pt]
	\item We propose five approximate quantum adders (three proposed adders with carryout, and two proposed adders without carryout) designed using approximate computing to reduce the depth and boost noise resilience.
	\item Out of the five designs, two are zero-depth parallel adders that do not require any quantum gates because they employ pass-through logic. The other three adders are constant depth parallel adders that compute the Sum output of the addition using single CNOT gates.
	\item We illustrate the superior noise-resilience of the proposed quantum approximate adders by conducting simulations based on IBM Qiskit noise models. 
	\item For various noise models used in this study: (i) Compared to exact quantum ripple carry adder without carryout the proposed approximate adders without carryout have improved fidelity ranging from 8.34\% to 219.22\%, and (ii) Compared to exact quantum ripple carry adder with carryout the proposed approximate adders with carryout have improved fidelity ranging from 8.23\% to 371\%.
	\item We also assess the error introduced by approximate computing in proposed adders by error metrics like NMED and error rate.
\end{itemize}
This work is organized as follows: Section \ref{Background} provides background on quantum noise sources and error metrics. Section \ref{Proposed Design} presents the proposed approximate adders with error deviation. Section \ref{Performance} compares the noise performance of proposed adders. Section \ref{Conclusion} discusses the results and concludes this work.

\section{Background}
\label{Background}
	\subsection{Quantum Gates}
\begin{figure}[h]
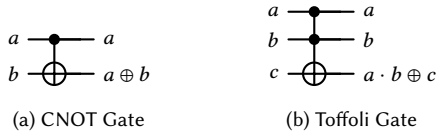

		\centering
		\small
		\begin{subfigure}[b]{0.2\textwidth}
			\centering
			\myCnot
			\caption{CNOT Gate}
			\label{fig:CNOT}
		\end{subfigure}
		\begin{subfigure}[b]{0.2\textwidth}
			\centering
			\myToffoli
			\caption{Toffoli Gate}
			\label{fig:Toffoli}
		\end{subfigure}	
		\caption{The CNOT and Toffoli gates.}
		\label{fig:qgates}
	\end{figure}

	The quantum gates used in this work are explained below.
	\begin{itemize} [leftmargin=+.2cm, topsep=2pt]
		\item CNOT Gate: CNOT or Feynman gate shown in Figure~\ref{fig:qgates}(a) helps in realizing XOR operation by mapping A, B to A, A$\oplus$B. 
		\item Toffoli Gate: Figure~\ref{fig:qgates}(b) shows the mapping of three input A, B, C to three outputs A, B, A$\cdot$B$\oplus$C. Toffoli gate, also known as CCNOT (double controlled NOT) gate, can achieve AND logical operation, and is the most resource intensive gate used in this work. 
	\end{itemize}

	\subsection{Quantum Computing Noise Sources}
	Noise is the main obstacle in the adoption and scaling of quantum computing \cite{nielsen2002quantum}. The prime source of quantum noise lies in the design and type of quantum computer. We focus on identifying unique noise sources and their impact on proposed designs by using IBM Qiskit \cite{qiskitModel} commands to inject uncertainty in simulations. Table \ref{table:ErrorDef} shows a compilation of noise models used in this work.
	\begin{table}[h]
		\caption{\textsc{Description of Noise Models}}
		\label{table:ErrorDef}
		\centering
		\addtolength{\tabcolsep}{-2pt}
		\begin{tabular}{p{0.17\linewidth}p{0.85\linewidth}}
			\hline
			\textbf{Model} & \textbf{Description} \\ \hline
			Thermal & Thermal relaxation error is characterized by time constants related by T1 (thermal relaxation) $\le$ T2 (dephasing). We use T1 = 50$\mu$s, T2 = 70$\mu$s \cite{qiskitModel}. \\
			Depolarizing & It is the probability that a completely mixed state I/2 replaces the state of qubit \cite{nielsen2002quantum}. We use an error probability of 0.005 for 1-qubit and 0.01 for 2-qubit gates. \\
			\begin{tabular}[c]{@{}l@{}}Amplitude\\ Damping\end{tabular} & This noise occurs due to energy dissipation during operations \cite{nielsen2002quantum}. We apply probability of 0.01 for 1-qubit gates.  \\
			\begin{tabular}[c]{@{}l@{}}Phase\\ Damping\end{tabular} & Phase damping involves the loss of information without energy loss \cite{nielsen2002quantum}. We use 0.01 probability for 1-qubit gates. \\
			SPAM & State Preparation And Measurement noise reflects the reliability. In state preparation, error probability of reset to 1 (0.04) is twice the error of reset to 0 (0.02), while measurement error is 0.1 \cite{paulnation2021} \\
			Bitflip & This noise source flips the qubit from 0 to 1 or vice versa, we use error probability for 1/2 qubit gates as 0.01 \cite{qiskitModel} \\
			Readout & P(n$\lvert$m) represents the probability of measuring n when correct output is m. We use P($1\lvert0$) = 0.05, P($0\lvert1$) = 0.1. \cite{qiskitModel} \\ \hline
		\end{tabular}
	\end{table}

	\subsection{Error Metrics}
	In this section, we explain the metrics used to measure the magnitude and frequency of the error introduced by the approximate adders. The most elementary metric is Error Distance (ED), which is the difference between the exact Sum ($S_{exact}$) and the approximate Sum ($S_{approx}$). We utilize ED to create Mean Error Distance (MED) by averaging over the total input combinations (N). In Equation \ref{equation:NMED}, normalization of MED by the maximum output across all inputs ($S_{max}$) reduces the sensitivity of NMED to outliers and variations in the distribution of error distances. NMED is particularly useful because it takes into account the scale and distribution of the error, allowing for fair comparison across different input configurations.
	\vspace{3mm}
	\begin{equation} \label{equation:ED}
		ED = \left | S_{exact} - S_{approx} \right |
	\end{equation}
	\vspace{2mm}
	\begin{equation} \label{equation:NMED}
		MED = \frac{\sum ED}{N} \qquad NMED = \frac{MED}{S_{max}}
	\end{equation}
	\vspace{1mm}
	
	Error rate (ER) is another metric used to evaluate the accuracy of an adder by measuring the proportion of incorrect results produced by it. ER is calculated by dividing the number of incorrect output ($N_{error}$) by the total input combinations (N), as shown in Equation \ref{equation:ER}. Adder with lower error rate will have errors limited to fewer input combinations, making them suitable for applications such as sensor data acquisition and image/audio filters which care for overall quality and discard outliers.
	\vspace{2mm}
	\begin{equation} \label{equation:ER}
		ER = \frac{N_{error}}{N}
	\end{equation}
	\vspace{1mm}	

\section{Proposed Approximate\\ Quantum Adders}
\label{Proposed Design}
\begin{figure}[h]
	\centering
	\vspace{3mm}
	\includegraphics[width=\columnwidth]{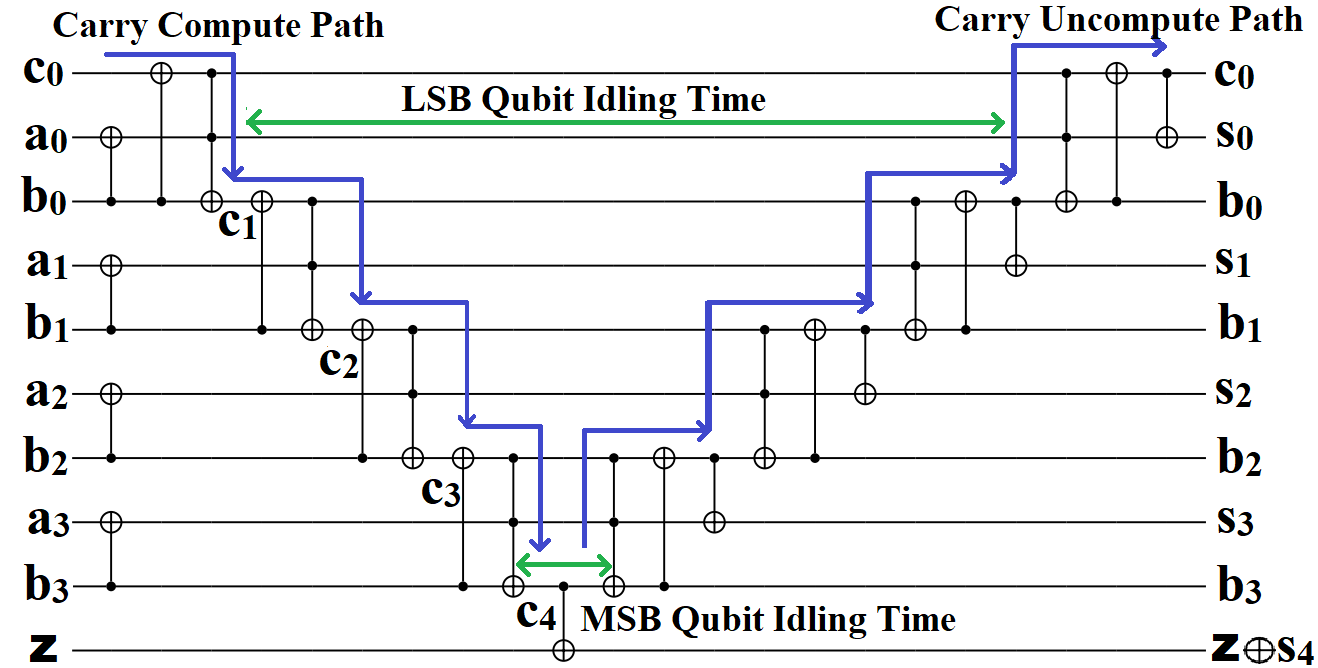}
	\caption{Exact quantum adder by Cuccaro et al \cite{cuccaro2004new}. The inputs are in range a\textsubscript{0} to a\textsubscript{4} and b\textsubscript{0} to b\textsubscript{4} while S\textsubscript{0} to S\textsubscript{4} represent Sum bits. The Carry generation path extends from c\textsubscript{0} to c\textsubscript{4}, creating higher depth and longer carrypath, increasing the noise susceptibility of carryout. LSB qubits spend much more time idling than MSB, making Sum vulnerable to noise \cite{jayashankar2022achieving}.} 
	\label{fig:exactAdderFig}
\end{figure}

In this section, we propose five approximate quantum adders to reduce the number of quantum gates and the depth to increase noise resilience. Figure \ref{fig:exactAdderFig} illustrates two root causes of noise in the design of existing quantum exact adders that we address in this work by utilizing approximate computing. First, we approximate the Sum output of the addition of two qubits using only its corresponding inputs. Our approximate adders operate in parallel, as there is no carry propagation during the addition operation. We do not generate intermediate carries and instead focus on creating the final carryout using only the MSB input qubits. These design techniques help generate all qubits of the Sum in parallel, avoiding the issue of idling qubits. In addition, the noise accumulated on carryout is lesser as there is no longer a carry path. Applying these design principles reduces the depth of proposed adders and increases their noise resilience.

Table \ref{table:Design} showcases the Sum and Carry logical equations for the exact quantum adders and the proposed quantum approximate adders. The proposed adders AQA1 and AQA2 are without Carry, whereas AQA3, AQA4, and AQA5 are with carryout. Table \ref{table:Design} also lists the CNOT and Toffoli gate count with their respective depth. The O(1) or constant depth complexity of proposed adders is superior to the O(n) or linear complexity of exact adders.

\newcommand{\sumExact}{$A \oplus B \oplus C_{in}$}
\newcommand{\carryExact}{$A \cdot B \oplus B \cdot C_{in} \oplus C_{in} \cdot A$}
\newcommand{\sumTPL}{$A \oplus B \oplus C_{i}$}
\newcommand{\carryTPL}{$A \cdot B \oplus B \cdot C_{i} \oplus C_{i} \cdot A$}
\newcommand{\sumApprox}{$A \oplus B$}

\begin{table*}[]
	\caption{\textsc{Design Characteristics of Exact and Proposed Approximate Adders}}
	\label{table:Design}
	\centering
	\begin{tblr}{
			row{2} = {c},
			column{3} = {c},
			column{4} = {c},
			column{5} = {c},
			cell{1}{1} = {r=2}{},
			cell{1}{2} = {r=2}{},
			cell{1}{3} = {r=2}{},
			cell{1}{4} = {r=2}{},
			cell{1}{5} = {r=2}{},
			cell{1}{6} = {c=2}{c},
			cell{1}{8} = {c=2}{c},
			cell{3}{1} = {r=3}{},
			cell{3}{6} = {c},
			cell{3}{7} = {c},
			cell{3}{8} = {c},
			cell{3}{9} = {c},
			cell{4}{6} = {c},
			cell{4}{7} = {c},
			cell{4}{8} = {c},
			cell{4}{9} = {c},
			cell{5}{6} = {c},
			cell{5}{7} = {c},
			cell{5}{8} = {c},
			cell{5}{9} = {c},
			cell{6}{1} = {r=2}{},
			cell{6}{6} = {c},
			cell{6}{7} = {c},
			cell{6}{8} = {c},
			cell{6}{9} = {c},
			cell{7}{6} = {c},
			cell{7}{7} = {c},
			cell{7}{8} = {c},
			cell{7}{9} = {c},
			cell{8}{1} = {r=3}{},
			cell{8}{6} = {c},
			cell{8}{7} = {c},
			cell{8}{8} = {c},
			cell{8}{9} = {c},
			cell{9}{6} = {c},
			cell{9}{7} = {c},
			cell{9}{8} = {c},
			cell{9}{9} = {c},
			cell{10}{6} = {c},
			cell{10}{7} = {c},
			cell{10}{8} = {c},
			cell{10}{9} = {c},
			vlines,
			hline{1,3,6,8,11,12} = {-}{},
			hline{2} = {6-9}{},
			hline{4-5,7,9-10} = {2-9}{},
		}
		\textbf{Adder Type} & \textbf{Name} & \textbf{Sum} & \textbf{Carry} & \textbf{Qubits} & \textbf{Depth} &  & \textbf{Count} & \\
		&  &  &  &  & \textbf{CNOT} & \textbf{Toffoli} & \textbf{CNOT} & \textbf{Toffoli}\\
		\textbf{Quantum Exact Adders} & CQA0\cite{cuccaro2004new} & \sumExact & N.A. & $2n + 1$ & $3n + 1$ & $2n$ & $4n$ & $2n$\\
		& CQA1\cite{cuccaro2004new} & \sumExact & \carryExact & $2n + 2$ & $3n + 2$ & $2n$ & $4n + 1$ & $2n$\\
		& TPL13\cite{thapliyal2013design} & \sumTPL & \carryTPL & $2n + 1$ & $3n - 2$ & $2n - 1$ & $5n - 5$ & $2n - 1$\\
		{\textbf{Proposed Adders }\\\textbf{Without Carryout}} & AQA1 & A & N.A. & $2n$ & 0 & 0 & 0 & 0\\
		& AQA2 & \sumApprox & N.A. & $2n$ & 1 & 0 & n & 0\\
		{\textbf{Proposed Adders}\\\textbf{With Carryout}} & AQA3 & A & $B_{n-1}$ & $2n$ & 0 & 0 & 0 & 0\\
		& AQA4 & \sumApprox & $B_{n-1}$ & $2n$ & 1 & 0 & n & 0\\
		& AQA5 & \sumApprox & $A_{n-1} \cdot B_{n-1}$ & $2n + 1$ & 1 & 1 & $n$ & 1
	\end{tblr}
\end{table*}

\subsection{Proposed Adders Without Carryout}
\begin{figure}[ht]
		\begin{subfigure}[t]{0.45\linewidth}
			\centering
			\includegraphics[width=45mm]{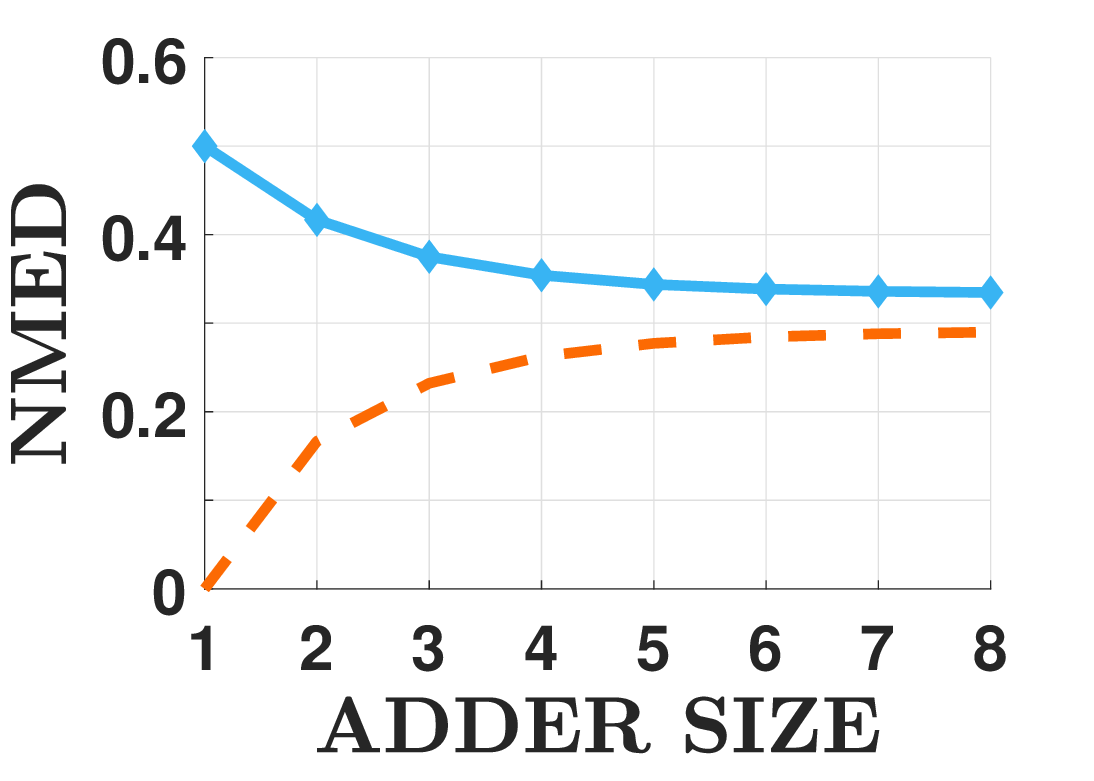}
			\caption{ }
			\label{fig:NMED1}
		\end{subfigure}
		\hfill
		\begin{subfigure}[t]{0.45\linewidth}
			\centering
			\includegraphics[width=45mm]{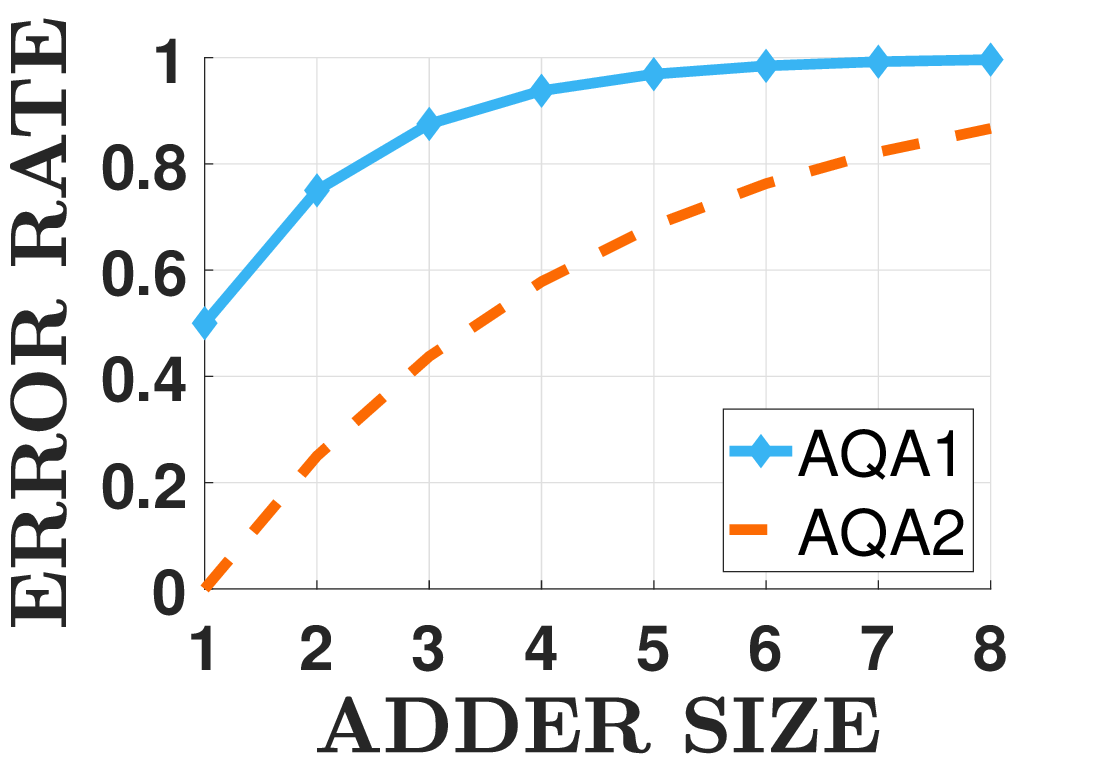}
			\caption{ }
			\label{fig:ER1}
		\end{subfigure}
		\caption{Error Metrics of Proposed Adders without Carryout for: (a) Normalized Mean Error Deviation. (b) Error Rate.}
		\label{fig:EMwithoutCarry}
	\end{figure}

	We first design adders without carryout for quantum applications that do not need an output carry. We only approximate the Sum for these adders, keeping the designs simple and with low depth.
	\begin{enumerate} [wide, labelwidth=!, labelindent=0pt, label=\textbf{(\arabic*)}]
		\item \textbf{Approximate Quantum Adder 1 (AQA1):}		
		The easiest way to approximate Sum is by using one of the inputs. In our first design AQA1, we map the input A to Sum and pass the input B unaltered, as shown in Equation \ref{aqa1}. As evident from Figure \ref{fig:AQA_NC}(a), the adder is zero depth, and utilizes no quantum gates.
		\begin{equation}
			s_i = a_i \;\;\;\; \text{if}\;0\leq{i}\leq{n-1}
			\label{aqa1}
		\end{equation}

		\item \textbf{Approximate Quantum Adder 2 (AQA2):}
		To create a better approximation than AQA1, we use logical-XOR of the two inputs as depicted in Equation \ref{aqa3}. This design is also without carryout, has single depth, and has n CNOT gates for n-qubit input, evident from Figure \ref{fig:AQA_NC}(b). As a result, AQA2 performs better than AQA1 regarding both NMED and error rate, as shown in Figure \ref{fig:EMwithoutCarry}. As evident, the proposed adder AQA2 scales better than AQA1 in the entire range upto eight qubits.
		\begin{equation}
			s_i = a_i \oplus b_i\;\;\;\; \text{if}\;0\leq{i}\leq{n-1}
			\label{aqa3}
		\end{equation}
	\end{enumerate}

		\newcommand{\figWidthA}{0.45\linewidth}
		\begin{figure}[h]
			\centering
			\begin{subfigure}[b]{0.45\linewidth}
				\centering
				\captionsetup{justification=centering}
				\passNoCarryTwo
				\caption{AQA1: Pass based\\Zero Depth  }
			\end{subfigure} 
			\begin{subfigure}[b]{0.5\linewidth}
				\centering
				\captionsetup{justification=centering}
				\xorNoCarryTwo
				\caption{AQA2: CNOT based\\Single Depth}
			\end{subfigure} 
			\vspace{3mm}
			\caption{Proposed 2-Qubit Adders without Carryout. a) AQA1 needs no hardware to pass input A to Sum. b) Single CNOT gate in AQA2 reduces error distance.}
			\label{fig:AQA_NC}
		\end{figure}

\subsection{Proposed Adders With Carryout}
\begin{figure}[ht]
	\begin{subfigure}[t]{0.45\linewidth}
		\centering
		\includegraphics[width=45mm]{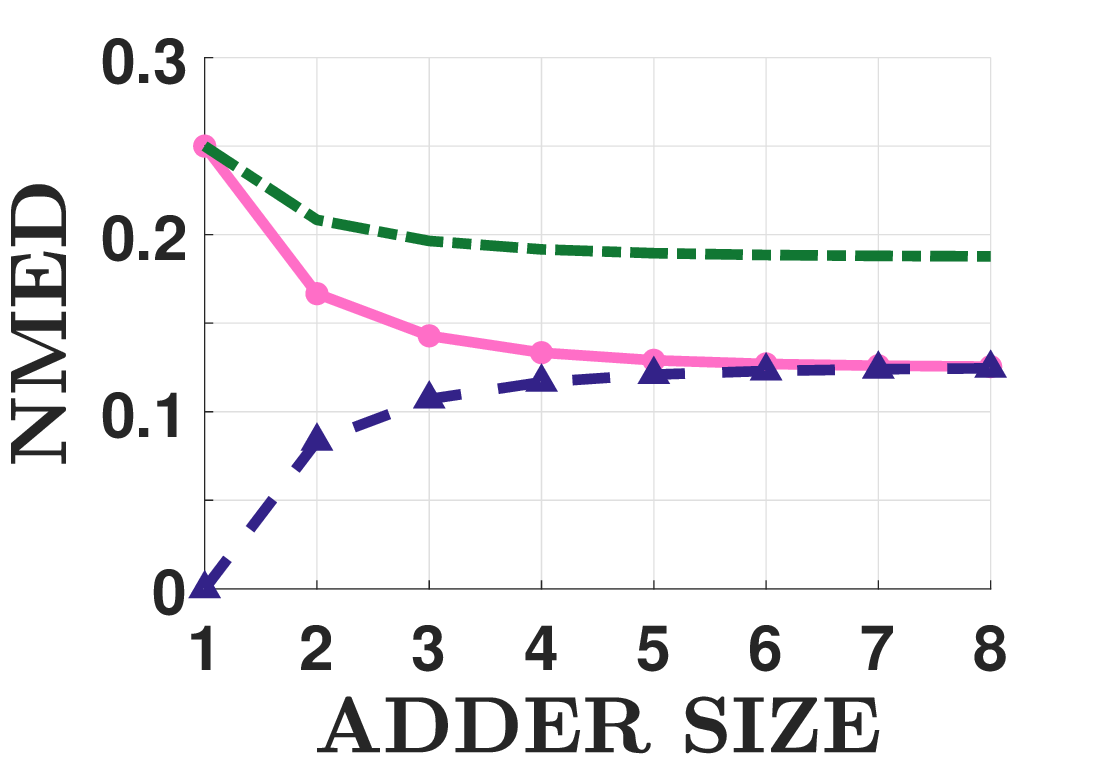}
		\caption{ }
		\label{fig:NMED2}
	\end{subfigure}
	\hfill
	\begin{subfigure}[t]{0.45\linewidth}
		\centering
		\includegraphics[width=45mm]{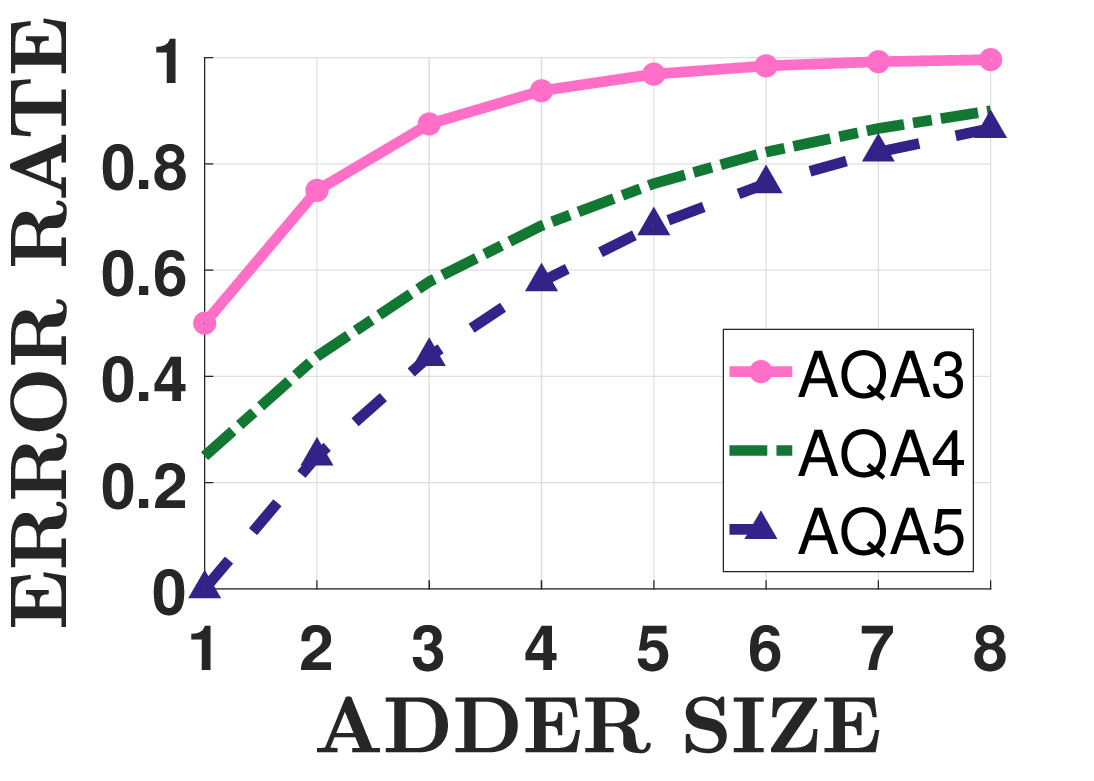}
		\caption{ }
		\label{fig:ER2}
	\end{subfigure}
	\caption{Error Metrics of Proposed Adders with Carryout for: (a) Normalized Mean Error Deviation (b) Error Rate.}
	\label{fig:EMwithCarry}
\end{figure}

	In several applications, the output carry generated from the addition of two inputs is required. Therefore, to design the approximate adder with carryout we explore design strategies to generate it with low depth and minimal gates encountered in the carry path. In the process, we also improve their error metrics.
	\begin{enumerate}[wide, labelwidth=!, labelindent=0pt, topsep=4pt, label=\textbf{(\arabic*)}]
		\item \textbf{{Approximate Quantum Adder 3 (AQA3):}}
		Our first design with an output carry AQA3, is derived from AQA1. The carryout is created from input B's MSB as shown in Equation \ref{aqa2}. Figure \ref{fig:AQA34}(a) illustrates that it has zero depth and zero gate count. Figure \ref{fig:EMwithCarry}(a) shows that inclusion of approximate carry improves the NMED from 0.35 to 0.13 for 4 qubit configuration, although the error rate remains same. Lower NMED value represents that the output will be closer to the exact output.
		\begin{equation} \label{aqa2}
			\begin{aligned} 		
				c_n=b_{n-1}\qquad 		  
				s_i=\begin{cases}
					a_i, & \text{if}\;0\leq{i}\leq{n-1} \\
					c_n, & \text{if}\;i=n 
				\end{cases} \\
			\end{aligned}
		\vspace{-3mm}
		\end{equation}
		\newcommand{\figWidthB}{0.15\textwidth}
		\begin{figure}[ht]
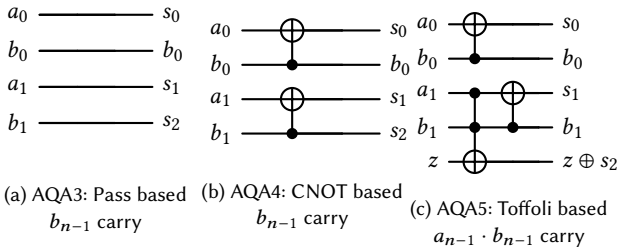

			\centering
			\begin{subfigure}[t]{\figWidthB}
				\centering
				\captionsetup{justification=centering}
				\passBCarryTwo
				\caption{AQA3: Pass based\\$b_{n-1}$ carry}
			\end{subfigure} 
			\begin{subfigure}[t]{\figWidthB}
				\centering
				\captionsetup{justification=centering}
				\xorBCarryTwo
				\caption{AQA4: CNOT based\\$b_{n-1}$ carry}
			\end{subfigure}
			\begin{subfigure}[t]{\figWidthB}
				\centering
				\captionsetup{justification=centering}
				\xorToffoliCarryTwo
				\caption{AQA5: Toffoli based\\$a_{n-1} \cdot b_{n-1}$ carry}
			\end{subfigure}
			\caption{Proposed 2-Qubit Adders With Carryout.}
			\label{fig:AQA34}
			\vspace{-2mm}
		\end{figure}
	
		\item \textbf{{Approximate Quantum Adder 4 (AQA4):}}
		Like in the case of AQA3, we also design AQA4 using AQA2 by reassigning the last qubit of the input B as carryout, as shown in Equation \ref{aqa4}. As observable from Figure \ref{fig:AQA34}(b), this approach helps ensure the depth remains constant with n CNOT gates for n inputs. The proposed adder AQA4 is worse than AQA3 regarding NMED but has a much superior error rate, as evident from Figure \ref{fig:EMwithCarry}.
		\begin{equation} \label{aqa4}
			\begin{aligned}
				c_n=b_{n-1}\qquad 				  
				s_i=\begin{cases}
					a_i \oplus b_i, & \text{if}\;0\leq{i}\leq{n-1} \\
					c_n, & \text{if}\;i=n 
				\end{cases} 
			\end{aligned}
		\end{equation}	
			
		\item \textbf{Approximate Quantum Adder 5 (AQA5):}
		We next propose approximate adder AQA5, which generates carryout as $a_{n-1} \cdot b_{n-1}$ to provide a better approximation of the carryout, where $a_{n-1}$ and $b_{n-1}$ are the input MSB's. We use a Toffoli gate for this purpose as shown in Figure \ref{fig:AQA34}(c). AQA5 provides output carry on a separate qubit for quantum applications that require cascading a larger quantum circuit after the quantum adder. AQA5 shields the cascaded circuit from input noise sources and provides a high fidelity carryout while passing an input for downstream utilization. Figure \ref{fig:EMwithCarry} shows that AQA5 is better than both AQA3 and AQA4 in terms of NMED and error rate.
		\begin{equation} \label{aqa5}
			\begin{aligned} 
				c_n=a_{n-1} \cdot b_{n-1}\qquad				  
				s_i=\begin{cases}
					a_i \oplus b_i, & \text{if}\;0\leq{i}\leq{n-1} \\
					c_n, & \text{if}\;i=n 
				\end{cases} \\
			\end{aligned}
		\end{equation}
		The proposed approximate adders reduce quantum gates to achieve zero or constant depth. Unlike exact adders, they generate parallel output on all Sum qubits, generating carryout using a shorter mechanism. Both these factors help improve the noise fidelity of proposed approximate adders, which we explore in Section \ref{Performance}.
\end{enumerate}

\section{Noise-Resilience Analysis\\ of Proposed Approximate\\ Quantum Adders}
\label{Performance}
	To evaluate the noise resilience of proposed approximate quantum adders, we utilize the probability of obtaining the intended output to measure noise fidelity. To arrive at the output probability, we simulate the adders in a 4-qubit configuration using IBM Qiskit for all 256 input configurations. Then, we divide the number of accurate results by the simulation count (256 in this case). By modifying only the noise model in qiskit simulations and leaving the rest unchanged, we perform the same process for each adder using the different noise models. 
	SPAM and readout noise depend on the measurement and input states making them proportional to the qubit count, impacting the exact and approximate adders in the same manner in our study. Due to this, we present the results of five noise models in Tables \ref{table:NoiseNoCarry} and \ref{table:NoiseCarry} for comparison. In the rest of this work, the average output probability is interchangeably referred to as accuracy (in percentage). To simplify the process of analyzing noise, we split the adders into two categories depending on the presence (or absence) of output carry. 
	
	\begin{table} [h]
		\caption{\textsc{Output Probability of Proposed 4-Qubit Adders without Carryout from Noise Simulations}}
		\label{table:NoiseNoCarry}
		\centering
		\resizebox{1.05\columnwidth}{!}{%
			\renewcommand{\arraystretch}{1.25}
			\begin{tabular}{|l|c|c|c|c|c|} 
				\hline
				\multirow{2}{*}{\textbf{Noise Type}} & \multirow{2}{*}{\textbf{CQA0\cite{cuccaro2004new}}} & \multirow{2}{*}{\textbf{AQA1}} & \textbf{Impr. \%} & \multirow{2}{*}{\textbf{AQA2}} & \textbf{Impr. \%} \\
				&  &  & \textbf{\textbf{\textbf{\textbf{w.r.t. CQA0}}}} &  & \textbf{\textbf{w.r.t. CQA0}} \\ 
				\hline
				\textbf{Thermal} & 0.712 & 0.951 & 33.57 & 0.935 & 31.32 \\ 
				\hline
				\textbf{Depolarizing} & 0.589 & 0.995 & 68.93 & 0.97 & 64.69 \\ 
				\hline
				\textbf{Phase} & 0.923 & 1.0 & 8.34 & 1.0 & 8.34 \\ 
				\hline
				\textbf{Amplitude} & 0.776 & 0.98 & 26.29 & 0.961 & 23.84 \\ 
				\hline
				\textbf{Bitflip} & 0.307 & 0.98 & 219.22 & 0.924 & 200.98 \\ 
				\hline
			\end{tabular}%
		}
	\end{table}
	
	\subsection{Comparison of noise resilience of proposed approximate adders without carryout}
	The adders in Table \ref{table:NoiseNoCarry} are the proposed adders without carryout simulated using noise conditions in 4-qubit configuration. 
	While comparing with exact adder without carry CQA0\cite{cuccaro2004new}: 
	\textbf{1. Thermal Noise}: AQA1 shows an improvement of 33.57\%, while AQA2 shows 31.32\%.
	\textbf{2. Depolarizing}: AQA1 shows 68.93\% improvement while AQA2 shows 4\% lower than it.
	\textbf{3. Phase Damping}: Both AQA1 and AQA2 show equal accuracy improvement of 8.24\%.
	\textbf{4. Amplitude Damping}: AQA1 shows about 3\% higher accuracy improvement than AQA2, which shows 23.84\%.
	\textbf{5. Bitflip}: AQA1 and AQA2 show an equally big improvement of 200\%.
	Our analysis demonstrates that the noise resilience of AQA2 is only marginally impacted by the inclusion of an extra CNOT gate.
	\begin{table}[h]
		\caption{\textsc{Output Probability of Proposed 4-Qubit Adders with Carryout from Noise Simulations}}
		\label{table:NoiseCarry}
		\centering
		\resizebox{\columnwidth}{!}{%
			\renewcommand{\arraystretch}{1.25}
			\begin{tabular}{|l|c|c|c|c|c|c|}
				\hline
				\textbf{Noise Type} & \textbf{CQA1\cite{cuccaro2004new}} & \textbf{TPL13\cite{thapliyal2013design}} & \textbf{AQA3} & \textbf{AQA4} &  \textbf{AQA5} \\ \hline
				\textbf{Thermal}      & 0.663 & 0.6956 & 0.953 & 0.926 & 0.904 \\ \hline
				\textbf{Depolarizing} & 0.576 & 0.6292 & 0.994 & 0.968 & 0.917 \\ \hline
				\textbf{Phase}        & 0.924 & 0.9408 & 1.0   & 1.0   & 0.99  \\ \hline
				\textbf{Amplitude}    & 0.772 & 0.8139 & 0.975 & 0.961 & 0.94  \\ \hline
				\textbf{Bitflip}      & 0.207 & 0.263  & 0.975 & 0.915 & 0.814 \\ \hline
			\end{tabular}%
		}
	\end{table}
	
	\subsection{Comparison of noise resilience of proposed approximate adders with carryout}
	For the proposed quantum adders with carryout listed in Table \ref{table:NoiseCarry}, we consider CQA1\cite{cuccaro2004new} as the baseline to compare the proposed adders and the more optimized TPL13\cite{thapliyal2013design}. While comparing with exact adders, Table \ref{table:NoiseCarryImpr} illustrates that:
	\textbf{1. Thermal Noise}: AQA3 shows improvement of 43.74\% with CQA1\cite{cuccaro2004new} and 37\% with TPL13\cite{thapliyal2013design}, while AQA4 shows 39.67\% and 33.12\%, and AQA5 shows 36.35\% and 29.96\% respectively.
	\textbf{2. Depolarizing}: All adders show high improvement. AQA3 showing 72.57\% and 57.98\% over CQA1\cite{cuccaro2004new} and TPL13\cite{thapliyal2013design} respectively, AQA4 showing 68.06\% and 53.85\%, while AQA5 showing 59.2\% and 45.74\% respectively.
	\textbf{3. Phase Damping}: Both AQA3 and AQA4 show improvement of 8.23\% with CQA1\cite{cuccaro2004new} and 6.29\% with TPL13\cite{thapliyal2013design}, with AQA5 behind by only 1\%.
	\textbf{4. Amplitude Damping}: AQA3 shows improvement of 26.3\% with CQA1\cite{cuccaro2004new} and 19.79\% with TPL13\cite{thapliyal2013design}. AQA4 and AQA5 follow closely with about 2\% and 4\% lower gain, respectively.
	\textbf{5. Bitflip}: All adders show great improvement. AQA3 leads with a wide margin of improvement by 371\% and 270\% over CQA1\cite{cuccaro2004new} and TPL13\cite{thapliyal2013design} respectively. AQA4 shows 342\% and 245\%, and AQA5 shows 293.24\% and 209.51\% over CQA1\cite{cuccaro2004new} and TPL13\cite{thapliyal2013design} respectively.
	AQA3 leads the proposed adders in noise fidelity improvement as it is zero depth, compared to AQA4 with single depth. AQA5 features a Toffoli gate which lowers error deviation but introduces a higher noise impact.
	
	\begin{table}[h]
		\caption{\textsc{Improvement Comparison of Proposed 4-Qubit Adders with Carryout with Existing Works}}
		\label{table:NoiseCarryImpr}
		\centering
		\resizebox{\columnwidth}{!}{%
			\renewcommand{\arraystretch}{1.25}
			\begin{tabular}{|l|c|c|c|c|c|c|} 
				\hline
				\textbf{Adder} & \multicolumn{1}{r|}{\textbf{\makecell{Impr.\%\\ w.r.t.}}}
				& \begin{sideways}\textbf{Thermal(\%)}\end{sideways} & \begin{sideways}\textbf{Depolarizing(\%)}\end{sideways} & \begin{sideways}\textbf{Phase(\%)}\end{sideways} & \begin{sideways}\textbf{Amplitude(\%)}\end{sideways} &  \begin{sideways}\textbf{Bitflip(\%)}\end{sideways} \\ 
				\hline
				\multirow{2}{*}{\textbf{AQA3}~} & \textbf{CQA1\cite{cuccaro2004new}} & 43.74 & 72.57 & 8.23 & 26.3 & 371.01 \\ 
				\cline{2-7}
				& \textbf{TPL13\cite{thapliyal2013design}}~ & 37 & 57.98 & 6.29 & 19.79 &  270.72 \\ 
				\hline
				\multirow{2}{*}{\textbf{AQA4}~} & \textbf{CQA1\cite{cuccaro2004new}} & 39.67 & 68.06 & 8.23 & 24.48 & 342.03 \\ 
				\cline{2-7}
				& \textbf{TPL13\cite{thapliyal2013design}}~ & 33.12 & 53.85 & 6.29 & 18.07 & 247.91 \\ 
				\hline
				\multirow{2}{*}{\textbf{AQA5}~} & \textbf{CQA1\cite{cuccaro2004new}} & 36.35 & 59.2 & 7.14 & 21.76 & 293.24 \\ 
				\cline{2-7}
				& \textbf{TPL13\cite{thapliyal2013design}}~ & 29.96 & 45.74 & 5.23 & 15.49 & 209.51 \\
				\hline
			\end{tabular}%
		}
	\end{table}

\section{Discussion and Conclusion}
\label{Conclusion}
We have proposed five approximate quantum adders in this work by optimizing the depth and gate count. We have shown that the proposed adders have better fidelity than the existing works by performing quantum simulations on the IBM qiskit platform, conducted using different noise models. Thermal and Bitflip noise hit exact adders the hardest since they have more gates, increasing the likelihood of output deviation. Depolarizing noise applies random pauli operation on the three-axis \cite{nielsen2002quantum}. Since this error needs to accumulate before affecting output, it impacts exact adders with higher depth and carry path for error accumulation. Phase damping noise, on the other hand, introduces a rotation along the z-axis \cite{nielsen2002quantum} but has minimal impact on the adders as our input lies on the x-axis. Amplitude damping noise is proportional to the idling time for qubit \cite{jayashankar2022achieving}, impacting the exact adders with ripple carry mechanism the severest. As the LSB in cuccaro exact adders must wait for the longest for the reverse computation, its output probability in amplitude damping noise model drops to 0.772. In contrast, the proposed approximate quantum adders are unscathed, as all qubits get processed in parallel. Finally, the proposed approximate quantum adders outperform the exact adders by a vast margin in both noise fidelity and gate count. We also study the error characteristics of the proposed approximate quantum adders using metrics such as NMED and error rate and establish the scalability of proposed approximate quantum adders. They can be very useful in applications such as quantum image processing which are inherently error tolerant and suited for approximate arithmetic. We conclude that the proposed quantum approximate adders can be very beneficial in the NISQ era, with our thorough error and noise analysis helping select an approximate adder suitable for the quantum application. \\[1pt]

\bibliographystyle{ACM-Reference-Format}
\bibliography{sample-base}


\end{document}